\documentclass[sigconf]{acmart} 
\AtBeginDocument{%
  \providecommand\BibTeX{{%
    \normalfont B\kern-0.5em{\scshape i\kern-0.25em b}\kern-0.8em\TeX}}}

\copyrightyear{2024} 
\acmYear{2024} 
\setcopyright{acmlicensed}\acmConference[UIST '24]{The 37th Annual ACM Symposium on User Interface Software and Technology}{October 13--16, 2024}{Pittsburgh, PA, USA}
\acmBooktitle{The 37th Annual ACM Symposium on User Interface Software and Technology (UIST '24), October 13--16, 2024, Pittsburgh, PA, USA}
\acmDOI{10.1145/3654777.3676377}
\acmISBN{979-8-4007-0628-8/24/10}

\usepackage{array}
\newcolumntype{L}[1]{>{\raggedright\let\newline\\\arraybackslash\hspace{0pt}}m{#1}}
\newcolumntype{C}[1]{>{\centering\let\newline\\\arraybackslash\hspace{0pt}}m{#1}}
\newcolumntype{R}[1]{>{\raggedleft\let\newline\\\arraybackslash\hspace{0pt}}m{#1}}

\usepackage{graphicx}
\usepackage{subcaption}
\usepackage{amsmath}
\usepackage{gensymb}

\newcommand{\TSD}{360$\degree$}

\begin{document}

\title{VirtualNexus: Enhancing 360-Degree Video AR/VR Collaboration with Environment Cutouts and Virtual Replicas}


\author{Xincheng Huang}
\authornote{Both authors contributed equally to this research.}
\affiliation{
  \institution{University of British Columbia}
  \city{Vancouver}
  \state{BC}
  \country{Canada}
}
\orcid{0000-0001-6923-6490}
\email{xchuang@cs.ubc.ca}

\author{Michael Yin}
\authornotemark[1]
\affiliation{
  \institution{University of British Columbia}
  \city{Vancouver}
  \state{BC}
  \country{Canada}
}
\orcid{0000-0003-1164-5229}
\email{jiyin@cs.ubc.ca}

\author{Ziyi Xia}
\affiliation{
  \institution{University of British Columbia}
  \city{Vancouver}
  \state{BC}
  \country{Canada}
}
\orcid{0000-0002-9856-9101}
\email{zxia0101@cs.ubc.ca}

\author{Robert Xiao}
\affiliation{
  \institution{University of British Columbia}
  \city{Vancouver}
  \state{BC}
  \country{Canada}
}
\orcid{0000-0003-4306-8825}
\email{brx@cs.ubc.ca}

\renewcommand{\shortauthors}{}

\begin{teaserfigure}
\begin{subfigure}[t]{0.24\textwidth}
    \includegraphics[width=\textwidth]{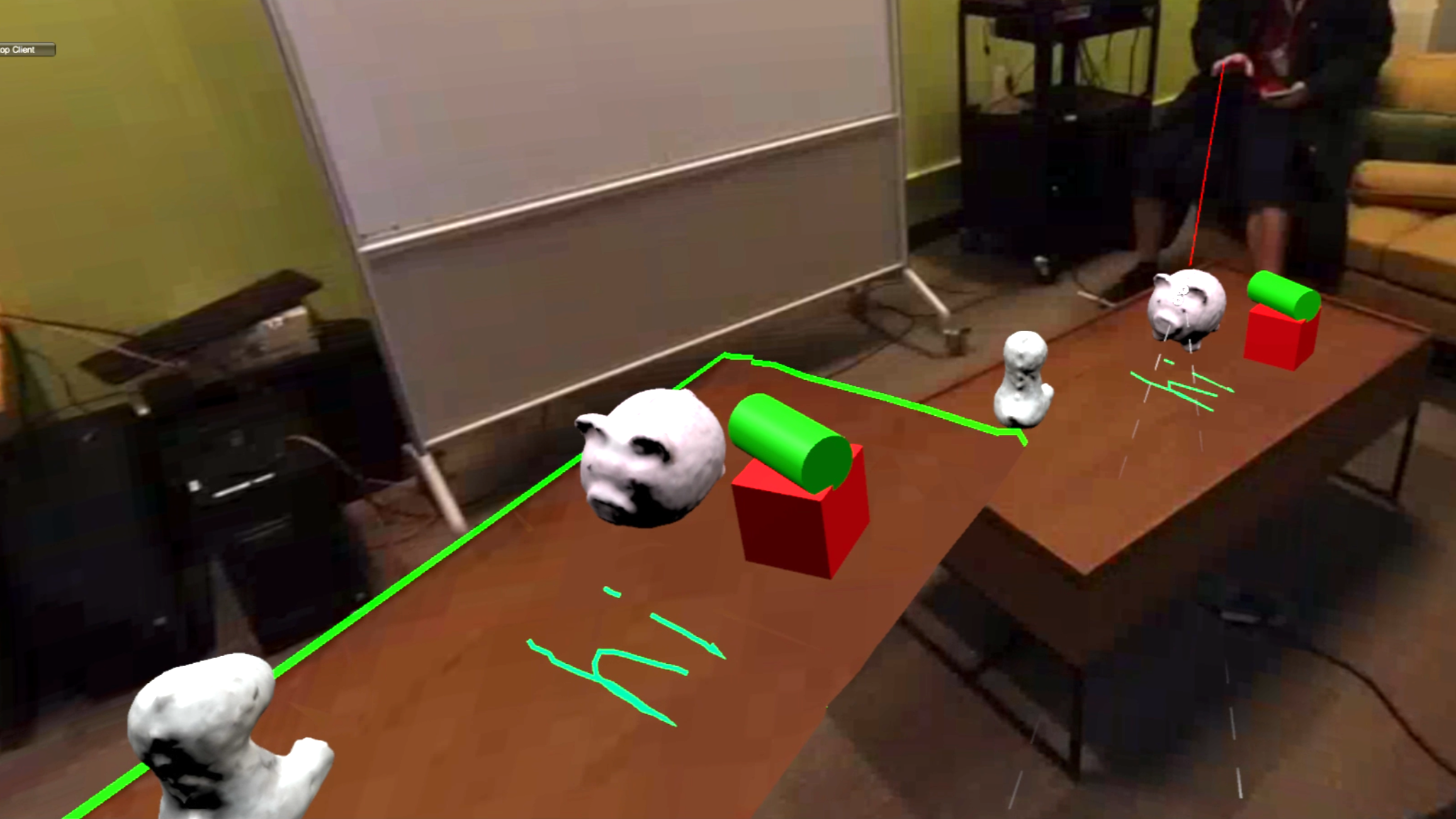}
    \caption{}
    \label{fig:teaser1}
\end{subfigure}
\hfill
\begin{subfigure}[t]{0.24\textwidth}
    \includegraphics[width=\textwidth]{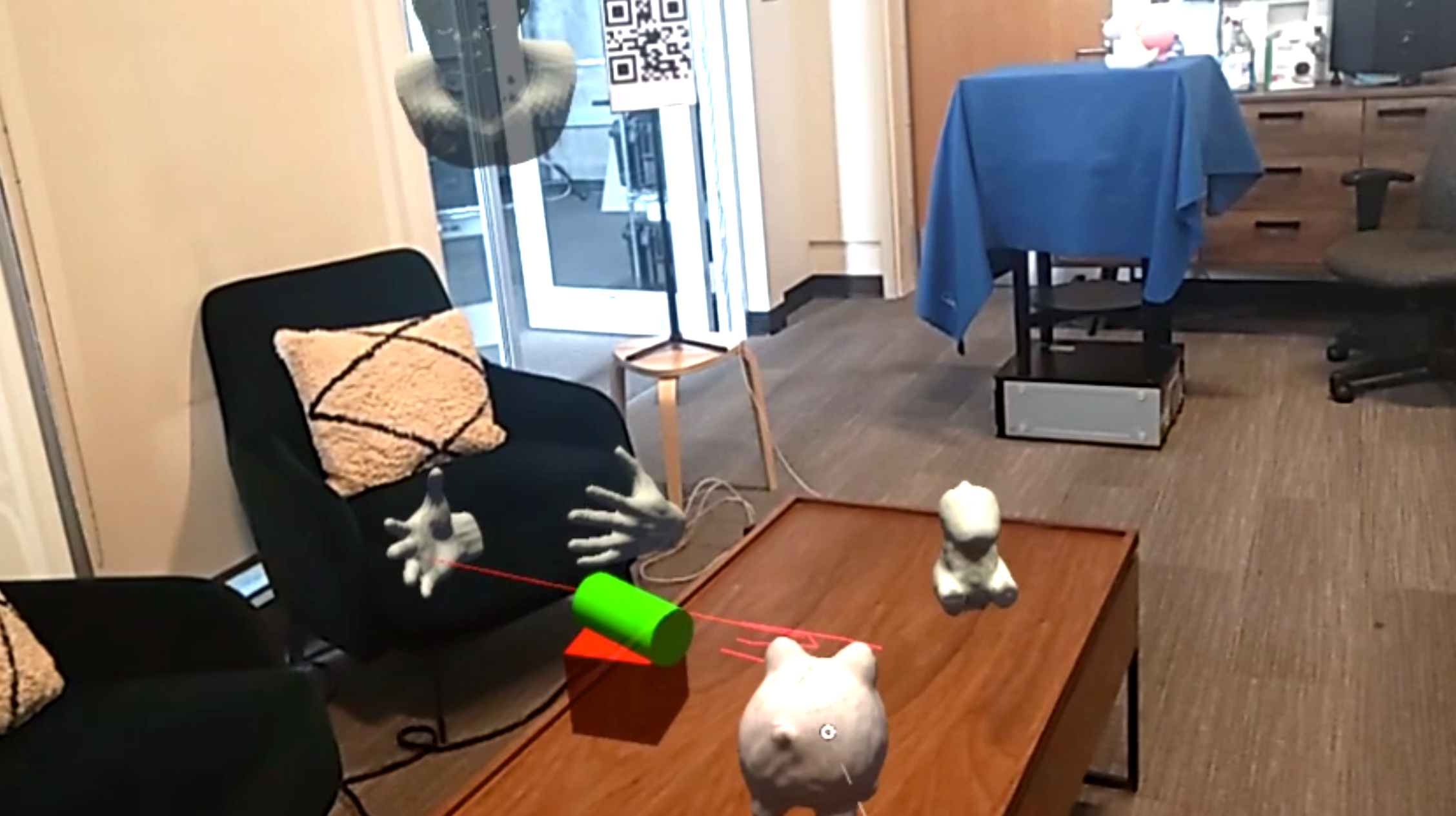}
    \caption{}
    \label{fig:teaser2}
\end{subfigure}
\hfill
\begin{subfigure}[t]{0.24\textwidth}
    \includegraphics[width=\textwidth]{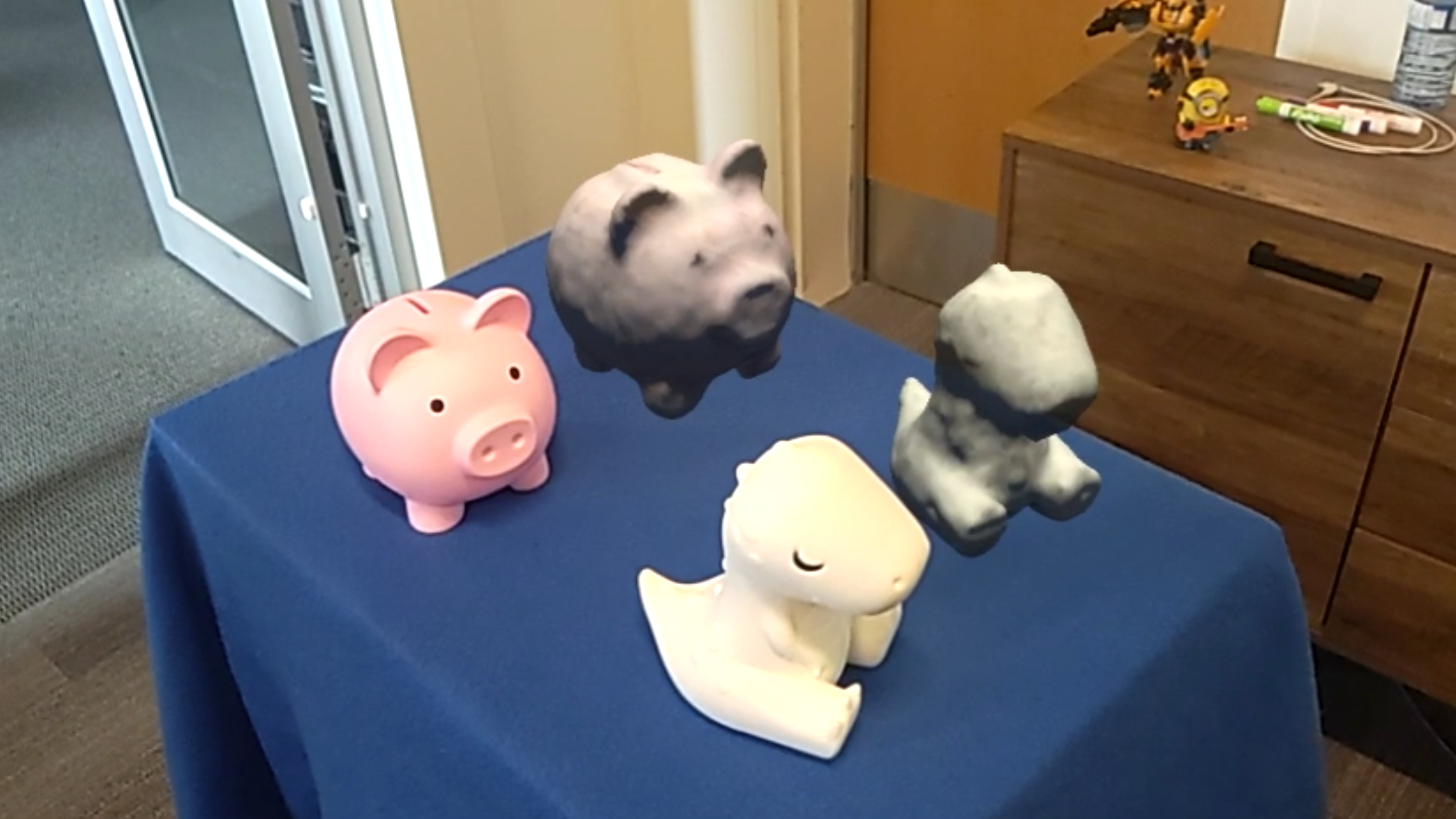}
    \caption{}
    \label{fig:teaser3}
\end{subfigure}
\hfill
\begin{subfigure}[t]{0.24\textwidth}
    \includegraphics[width=\textwidth]{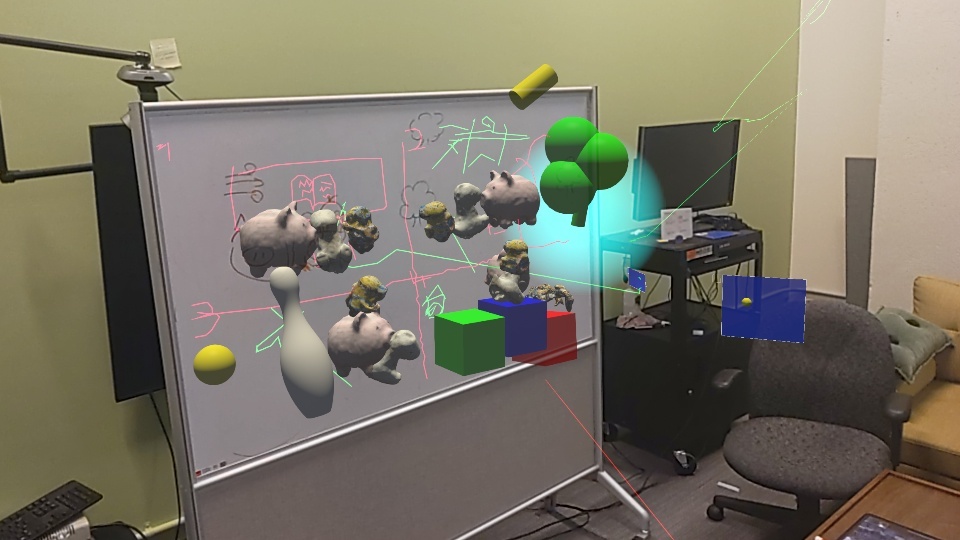}
    \caption{}
    \label{fig:teaser4}
\end{subfigure}
\caption{\textit{VirtualNexus} enhances \TSD{} video AR/VR collaboration with environment cutouts and virtual replicas. In (a), the VR user is telepresent in the AR user's physical environment. To have a close-up view of the desk, the VR user can create a manipulable \textit{environment cutout} that they have dragged closer. Simultaneously in (b), the AR user sees the VR user's avatar closer to the desk from the camera's position. Virtual annotations and objects are positionally synchronized across the cutout, the desk in the \TSD{} scene, and the physical desk. \textit{VirtualNexus} additionally implements ad-hoc 3D virtual replica creation from Instant-NGP~\cite{mueller2022instant}. In (c) we showcase virtual replicas of a pig and a dinosaur with their original physical copies. In (d) we showcase an example storyboard participants created with \textit{VirtualNexus} in our study (from AR user's perspective).}
\Description{This is the teaser figure. In (a) we see a VR user is telepresent in a lab space. The VR user is interacting with an environment cutout (i.e., a synchronized clone of the desk surface) that is pulled closer to the VR user. The VR user has placed a virtual replica of a dinosaur, a virtual replica of a pig, a green cylinder and a red cube on the desk. The VR user has also annotated “hi” on the cloned desk cutout. All the virtual objects and annotations on the cutout are spatial-accurately synchronized to the original desk with their original copy in the 360-degree scene. Simultaneously in (b), we see the perspective of the AR user, who is in the physical lab space. Because in (a) the VR user has pulled the cutout closer to them, and in (b), the AR user sees the virtual avatar of the AR user moved closer to the actual physical desk. The same virtual objects also spatially show up in (b) in the AR user’s environment. In (c), we demonstrate the ad-hoc 3D virtual replica creation with our system. We see the virtual replicas of a pig and a dinosaur next to their original physical copies. In (d) we showcase an example storyboard participants created in our study from the AR user's perspective. They have used various shapes, objects, annotations, and a cutout in order to create a storyboard for the provided narrative.}
\label{fig:teaser}
\end{teaserfigure}

\begin{abstract}
  Asymmetric AR/VR collaboration systems bring a remote VR user to a local AR user’s physical environment, allowing them to communicate and work within a shared virtual/physical space. Such systems often display the remote environment through 3D reconstructions or \TSD{} videos. While \TSD{} cameras stream an environment in higher quality, they lack spatial information, making them less interactable. We present \textit{VirtualNexus}, an AR/VR collaboration system that enhances \TSD{} video AR/VR collaboration with \textit{environment cutouts} and \textit{virtual replicas}. VR users can define cutouts of the remote environment to interact with as a world-in-miniature, and their interactions are synchronized to the local AR perspective. Furthermore, AR users can rapidly scan and share 3D virtual replicas of physical objects using neural rendering. We demonstrated our system’s utility through 3 example applications and evaluated our system in a dyadic usability test. \textit{VirtualNexus} extends the interaction space of \TSD{} telepresence systems, offering improved physical presence, versatility, and clarity in interactions. 
\end{abstract}





\maketitle

\section{Introduction}
Asymmetric remote AR/VR collaboration systems allow a remote VR user to be telepresent in a local AR user's physical environment~\cite{teoMixedRealityRemote2019a, heCollaboVRReconfigurableFramework2020a, Thoravi_Kumaravel2019-Loki}, allowing them to communicate and work effectively within a shared virtual/physical space. Such systems usually display the physical environment to the remote VR user through 3D reconstructions (e.g., textured spatial meshes~\cite{teoExploringInteractionTechniques2020a, teoTechniqueMixedReality2019a}, point clouds~\cite{Thoravi_Kumaravel2019-Loki, Orts2016Holoportation, teoMixedRealityRemote2019a}), or \TSD{} videos. Typically, the decision between the two options induces a trade-off --- while \TSD{} videos stream in higher quality compared to 3D reconstructions, they hinder efficient bi-directional interaction as they lack 3D spatial (depth) information. Furthermore, virtual objects can only float in front of the \TSD{} video (instead of physically reacting with a 3D scene reconstruction), breaking the illusion of being physically present. 

To address the lack of spatial context, contemporary \TSD{} systems~\cite{Piumsomboon2019ShoulderGiants, JonesVROOM2021, JackInKasahara2015} have explored mobile locomotion for the \TSD{} camera. However, a locomotive camera may cause simulator sickness and is less feasible for regular users. In regards to the issue of virtual object reference and interaction, prior \TSD{} systems have enhanced functionality in collaboration through additional non-verbal cues such as gazes, gestures, ray pointers, and annotations~\cite{Piumsomboon2017ExploreEnhancement, teoExploringInteractionTechniques2020a, teoTechniqueMixedReality2019a, lee2018user}. However, similar enhancements have not been extended to object manipulation; it is also challenging to incorporate physical objects of the \TSD{} environment into the collaboration. Thus, a clear gap emerges - how can we retain the high visual fidelity of a \TSD{} display while extending the collaborative interaction with spatial manipulation, within the environment and with the virtual objects (more akin to 3D reconstruction)? Prior work has explored combining \TSD{} videos and 3D reconstructions in telepresence and remote collaboration~\cite{gao2021user, teoTechniqueMixedReality2019a, young2020mobileportation}; however, past research switches between these two views instead of harnessing their merits simultaneously.

We present \textit{VirtualNexus}, a system that augments spatial interactivity in standard \TSD{} video remote AR/VR collaboration using \textit{environment cutouts} and \textit{virtual replicas}. Environmental cutouts are a feature that allows a remote VR user to cut out a part of the \TSD{} environment as a live textured mesh. The users can pull the environment cutout closer as a World in Miniature (WiM), bringing the environment within reach and offering precise control. Changes a user makes in the environment cutout synchronize to the original \TSD{} video and overlay on the AR user's view of the physical environment. We further implemented ad-hoc 3D virtual replica creation with Instant-NGP~\cite{mueller2022instant}, which allows the local AR user to scan a physical object and obtain a shared virtual replica within 1--3 minutes, further bridging the physical and virtual environments. We demonstrated the utility of these novel features through three application scenarios and evaluated our system in a user study. \textit{VirtualNexus} is lightweight as we only require the use of an off-the-shelf \TSD{} camera, AR and VR HMDs, and a consumer-grade computer to act as the server. We found that \textit{VirtualNexus} extends the interaction space of \TSD{} telepresence systems with enhanced physical presence, versatility, and clarity in interactions. 

\section{Related Work}
Telepresence immersively brings a remote guest to a local user's physical environment~\cite{360VideoChatTang2017, Opt360DeliveryQian2016, FlareQian2018, Huang360VR2023}. It has been a longstanding area of research, especially in the context of AR/VR remote interaction~\cite{teoExploringInteractionTechniques2020a, teoMixedRealityRemote2019a, Huang2024SurfShare, Irlitti2023MRTelepresence, lee2018user}. To display the physical environment to the remote VR user, prior research has explored 3D reconstruction (i.e., textured spatial meshes~\cite{Piumsomboon2017ExploreEnhancement, 360DIVArtois2023} or point clouds~\cite{teoMixedRealityRemote2019a, Orts2016Holoportation}) and \TSD{} videos~\cite{JonesVROOM2021, Piumsomboon2019ShoulderGiants, liOmniGlobeVRCollaborative3602020a, LeeTeo2017MRCollab}. As 3D reconstructions are themselves virtual objects in VR, they have richer interactive potential than \TSD{} videos. It is easier for users to move around and augment a 3D reconstruction in a virtual world~\cite{Thoravi_Kumaravel2019-Loki, Orts2016Holoportation}. However, compared to \TSD{} videos, real-time 3D reconstruction typically has lower quality, and it suffers from holes and occlusions. Holoportation~\cite{Orts2016Holoportation} implements a pipeline that can stream high-quality full-scene reconstruction in real-time, but it requires high-end sensors, computing, and network infrastructure. In comparison, telepresence with \TSD{} video cost-effectively provides higher quality (commodity 6K \TSD{} cameras are around \$500) and thus better presence and immersion~\cite{presence1997Slater, PresenceQuestion1998Witmer, lee2018user}. Nevertheless, \TSD{} videos are essentially a texture rendered on a spherical screen. Therefore, it is more challenging to incorporate common AR/VR interactive modalities in \TSD{} telepresence. 

\subsection{Combining \TSD{} Video and 3D reconstructions}
Given the respective merits of \TSD{} video and 3D reconstructions, prior work has explored combining the two in remote AR/VR collaboration. Teo et al. and Gao et al. proposed toggling between the modes of using 3D reconstruction or \TSD{} video ~\cite{teoMixedRealityRemote2019a, gao2021user}. However, the need to switch between two different media prevents simultaneously harnessing the merit of both. The authors also reported that frequently switching between perspectives and interactive modalities offered by different modes is challenging to adjust to. Young et al. extended this work, providing seamless transitions based on distance between users instead ~\cite{young2020mobileportation}. Teo et al. also proposed follow-up works~\cite{teoExploringInteractionTechniques2020a, teoTechniqueMixedReality2019a} that can insert \TSD{} panorama as bubbles into 3D reconstructions. However, the 3D reconstruction in the proposed system has a static texture and is mostly used as context. Although users may update the 3D reconstruction's context with newly captured \TSD{} images, they rely mostly on the live \TSD{} video mode~\cite{teoExploringInteractionTechniques2020a} or live \TSD{} insertion~\cite{teoTechniqueMixedReality2019a} for real-time interaction. In our work, both content delivered through \TSD{} and 3D reconstruction are live. We simultaneously provide a live \TSD{} environment and live environment cutouts (spatial mesh textured with live video texture). We additionally provide enhanced interactivity with virtual objects and replicas. Thus, we now review common interactive requirements in AR/VR remote collaboration and how they apply to \TSD{} video.

\subsection{Interactivity in \TSD{} Video Telepresence}
To enhance the presence of the remote guest and the effectiveness of AR/VR remote collaboration, prior research has explored a variety of interactive modalities, and we review them as follows.

\subsubsection{Access and Exploring a \TSD{} Scene}\label{sec: rel_work_explore}
It is straightforward to allow users to move and explore the remote environment in a 3D reconstruction. However, the same task is more challenging for \TSD{} video telepresence as the remote users always take the perspective of the \TSD{} camera. With a stationary camera, users can only access farther regions of the scene with far manipulation (e.g., far hand ray), reducing the precision of control. Prior research has proposed having the local user move the \TSD{} camera in the physical space~\cite{JackInKasahara2015, teoExploringInteractionTechniques2020a, teoMixedRealityRemote2019a, teoTechniqueMixedReality2019a, Piumsomboon2019ShoulderGiants} by mounting a \TSD{} camera to the local user, synchronizing the perspective of the local user and the remote guest. However, such an approach leads to an inconsistency between the remote user's physical and perceived motion, which could lead to simulator sickness in VR~\cite{Piumsomboon2019ShoulderGiants, SimulatorSick2021Hirzle}. More importantly, transferring the perspective control to the local user diminishes the remote user's freedom to explore the space, which could impair more comprehensive collaborative tasks (e.g., prototyping, gaming, and entertainment, tasks with divided labour). Alternatively, VROOM~\cite{JonesVROOM2021} mounts a \TSD{} camera on a locomotive robotic agent remotely controlled by the remote user. However, using a robotic agent is too bulky and costly for regular users. 

\subsubsection{Worlds in Miniature}
Worlds in Miniature (WiM)~\cite{Danyluk2021WiM} is a miniaturized representation of an entire or part of a physical or virtual world. The most common use of WiMs is navigation~\cite{Kalkusch2002IndoorWIM, Mulloni2012IndoorNav}, but prior research has extended their capability to manipulate virtual environments~\cite{Danyluk2021WiM, Stoakley1995VRWIM, Coffey2011WIM}. Similar to manipulating a Voodoo doll~\cite{Pierce1999Voodoo}, synchronizing a user's inputs to a WiM with the larger world allows them to manipulate regions that are out of their reach. In an AR collaboration context, Yu et al. explore the idea of duplicated reality \cite{yu2022duplicated}. Their system creates a WiM (digital twin) that reconstructs a volume of the physical world, however, they rely on sensors in a small spatial region, limiting flexibility. Overall, using a WiM as an interactive technique in telepresence and remote collaboration has not been widely explored.  

\subsubsection{Reference and Augmentation}
In remote mixed-reality collaboration, users often augment the shared space with pointers, virtual annotations, and virtual objects so they can better communicate ideas and collaborate~\cite{Piumsomboon2017ExploreEnhancement, Seungwon2019EvaluatingCues}. The ability to reference and augment the virtual world enriches the task and collaboration space of remote communication~\cite{Buxton2009-vp} and facilitates group awareness~\cite{Gutwin2002WorkspaceAwareness}. Prior research has enhanced \TSD{} video collaboration with the use of gaze, ray pointers, and virtual annotations in \TSD{} videos~\cite{teoExploringInteractionTechniques2020a, teoMixedRealityRemote2019a, teoTechniqueMixedReality2019a, Teo2018handgestureannotation}. However, enhancing virtual object manipulation in \TSD{} remote collaboration has not been well explored. While it is common to have virtual objects react to the physical environment with collision and physics in mixed reality, \TSD{} videos lack spatial information to provide the same physicality (e.g., virtual objects float in front of the video, instead of lying on a physical surface), hindering the sense of being physically present for the remote user. Rhee et al.~\cite{Rhee2020ARVRTeleportation} incorporated synchronized ray pointers and virtual objects in remote collaboration. However, they took a graphical approach and focused on naturally blending virtual objects with the \TSD{} video using an image-based lighting technique for \TSD{} videos~\cite{Rhee2017MR360}. We take a physics approach: virtual objects are rendered on the \TSD{} video, but physically react to an embedded 3D reconstruction.

\subsection{Virtual Replicas and Neural Radiance Fields}
It is challenging to provide remote users access to the physical environment they are telepresent in. Recent research has taken mechanical and robotic approaches, allowing remote users to move physical objects in the local user's space with mini-robots~\cite{Ihara2023Holobots} or deformable interfaces~\cite{Leithinger2013Inform, Leithinger2014PhysicalTelepresence}. However, such methods usually have a limited area of operation (e.g., a delegated platform like a desk) and introduce additional hardware overhead. An alternative approach is to provide indirect physical access through virtual replicas~\cite{Oda2015VirRep, Elvezio2017VirtualRep}. However, most prior work requires virtual replicas to be created in advance with CAD tools~\cite{Zhang2022gesturevirtualreplica, Oda2015VirRep, Elvezio2017VirtualRep, Wang2023Behere, Wang20213DGAM} or only support creating from 2D contents or sketches~\cite{Huang2024SurfShare, Hu2023ThingShare, heCollaboVRReconfigurableFramework2020a}. While depth-based methods such as Kinect-Fusion~\cite{Izadi2011KinectFusion} can quickly reconstruct an object or a scene, more recently, Neural Radiance Fields (NeRF)~\cite{kangle2021dsnerf, Barron2021NeRF, Mildenhall2021NeRF, mueller2022instant} allow object and scene reconstruction with high quality. Notably, Instant-NGP~\cite{mueller2022instant} drastically reduces the training time of NeRF, making it feasible to reconstruct individual objects within seconds or minutes. In our work, we incorporate virtual replica creation with Instant-NGP into our collaborative telepresence system.

\section{Interactive Design and Concepts}
By distilling the requirements and gaps from related work, here we propose concepts and designs for \textit{VirtualNexus}.

\subsection{Preserving Spatial Physicality: Embedded 3D Reconstruction}
\TSD{} VR telepresence allows a user to explore a remote space omnidirectionally with immersion. However, regular \TSD{} videos lack spatial information to allow users to virtually interact with physics and collision (e.g., draw annotations on a wall, and bounce a virtual object on a desk), thus reducing the sense of being physically present. To solve this, we propose to align a spatial reconstruction with the \TSD{} Video. While we render virtual objects with the \TSD{} video, they behave like reacting with an actual physical environment when users manipulate them. The aligned spatial reconstruction should be transparent to preserve the higher reality of the \TSD{} video. As most state-of-the-art AR headsets maintain a spatial map behind the scene, the process of creating and aligning a 3D reconstruction should be seamless and hidden from the users.

\begin{figure*}[h]
  \centering
  \includegraphics[width=0.9\linewidth]{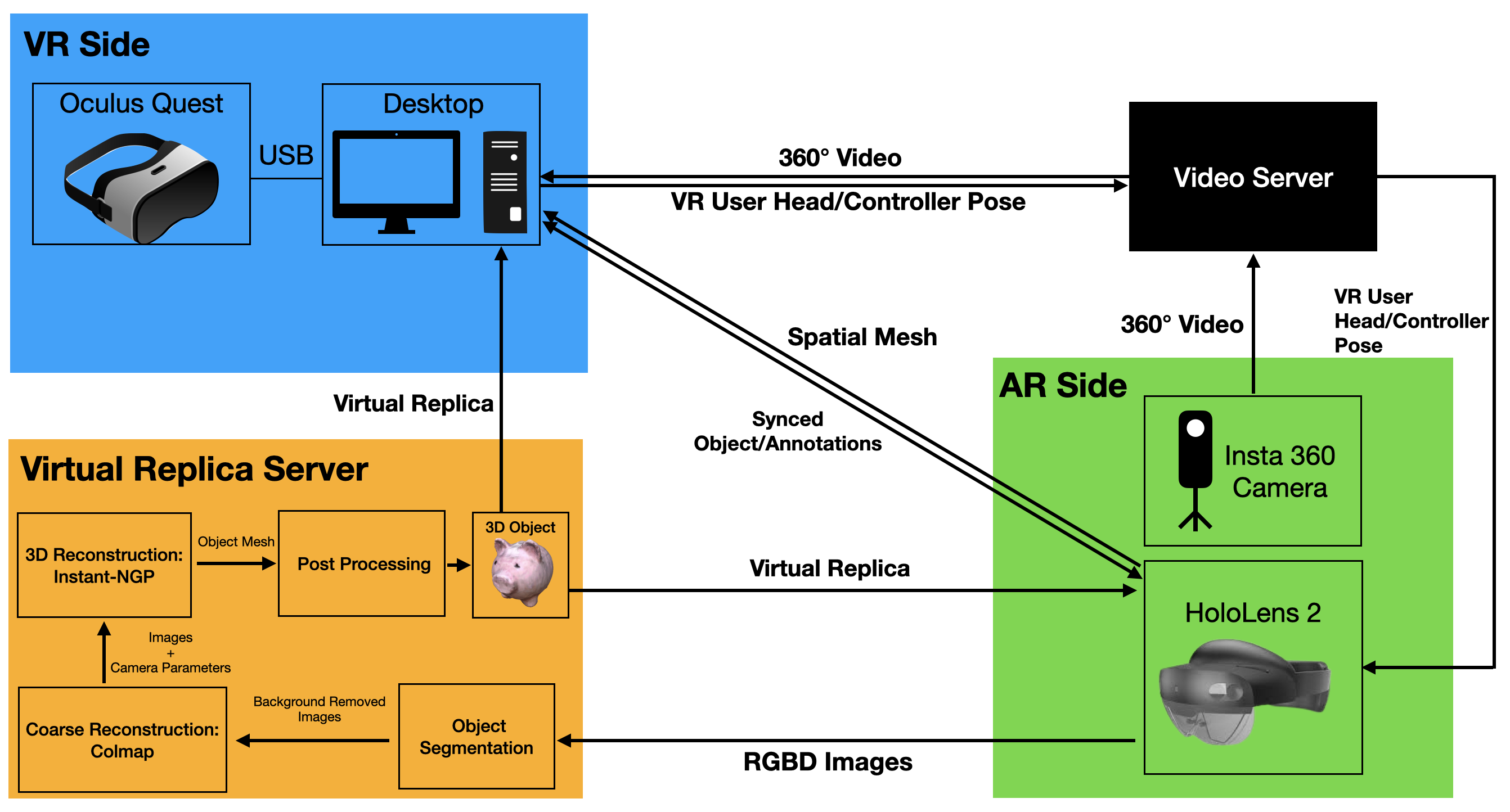}
  \caption{The system has four major components: VR side, AR side, video server, and virtual replica server. The VR side receives \TSD{} video from the AR side via the video server. The AR side shares synced objects and annotations with the VR side and sends scanned RGBD images to the virtual replica server, which creates virtual replicas and sends it to both the AR and VR sides.}
  \Description{Top left shows the VR side components, which consist of a desktop and a connected Oculus Quest 2 VR headset. The top right shows the video server. The bottom right shows the AR side components, which consist of the Insta360 camera and HoloLens 2. The bottom left shows the virtual replica server, which contains the pipeline for generating virtual replicas.}
  \label{fig:sys_arch}
\end{figure*} 

\subsection{Enhancing Access to Environments: Interactable Environment Cutouts}

In \TSD{} telepresence, with a regular stationary \TSD{} camera setup, users can only rely on far-hand manipulation (e.g., dragging an object with a long ray pointer) to access faraway regions in the scene, precluding precise interactions. Therefore, we introduce the concept of \textit{environment cutouts}, allowing the remote VR user to create a ``slice'' of the \TSD{} environment that can be interacted with at a different scale or position. For example, the VR user can select a part of the real world to make a copy, optionally scale it down (similar to a miniature diorama), pull it closer, and interact with this cutout (for example, placing virtual objects on this smaller world) while any such interactions are also reflected on the original world location. While the remote VR user can use the ray pointer to access farther objects, the ability to pull an environment cutout closer allows users to harness near interactions (e.g., grab, near draw) that have a higher precision. To convey the intention of the VR user to the AR user, the AR person will see the VR user's avatar moving toward the physical counterpart of the cutout as they pull an environment cutout (e.g., the VR person pulling a whiteboard closer is rendered as them moving toward it).

\subsection{From Reality to Virtual: Ad-hoc Creation of 3D Virtual Replicas}
In immersive environments, virtual replicas are useful props for referring to objects, conveying ideas, and prototyping rapidly~\cite{Elvezio2017VirtualRep, Zhang2022gesturevirtualreplica}. \textit{VirtualNexus} enables ad-hoc creation of 3D virtual replicas in remote AR/VR collaboration. The AR user can conveniently set an object on a platform, scan around it, and obtain a shared virtual replica. \textit{VirtualNexus} additionally stores the scanned virtual replica, enabling a ``scan once, create many'' experience. 

\subsection{Spatially Aligned and Synced Collaboration}
Co-location in a spatially aligned and synchronized environment is fundamental for remote AR/VR collaboration systems. Therefore, \textit{VirtualNexus} offers synchronized ray pointers, annotations, and shared virtual objects, which are essential elements to maintain group awareness~\cite{Buxton2009-vp, Gutwin2002WorkspaceAwareness}. For coherence, these features also adapt to the aforementioned system design: 1) annotations and virtual objects are able to collide and physically interact with the hidden spatial reconstructions, and 2) the environment cutout maintains a cloned copy of annotations and virtual objects that are synced with the original \TSD{} environment and the AR physical environment.

\section{VirtualNexus}
\label{sec: sys_impl}

\subsection{System Architecture and Apparatus}\label{sec: sys_arch}
We implemented \textit{VirtualNexus} (Fig. \ref{fig:sys_arch}) using Unity 2021.3.20f1, which can be configured as either a VR or AR application. \textit{VirtualNexus} uses Microsoft HoloLens 2\footnote{\url{https://www.microsoft.com/en-us/hololens}} for AR and Meta Oculus Quest 2\footnote{\url{https://www.meta.com/ca/quest/products/quest-2/}} for VR. An Insta360 X3\footnote{\url{https://www.insta360.com/product/insta360-x3}} \TSD{} camera omnidirectionally streams the local user's environment at 5.6K resolution and 30fps to the remote VR side. For efficient \TSD{} video streaming, we re-implemented a foveated video compression pipeline introduced by prior work \cite{Huang360VR2023} on a desktop machine with an Intel Core i7-9700K 3.6GHz CPU, 32GB RAM, and an NVIDIA GeForce RTX 2060. QR codes on the front and back of \TSD{} camera's tripod serve as the spatial anchors, aligning the VR world's origin with the \TSD{} camera's lenses. The local AR user sees a virtual avatar overlaid on the \TSD{} camera with synchronized head and hand poses of the remote VR user. Finally, \textit{VirtualNexus} runs virtual replica processing and VR-side rendering on the same machine, which has an Intel Core i9-12900KF 3.2GHz CPU, 64GB memory, and an NVIDIA GeForce RTX 4090 GPU. The AR user can scan a physical object and send the resulting photos to the virtual replica creation server. The server pre-processes the images, reconstructs a virtual replica with Instant-NGP, and sends it back to both AR and VR as shared virtual objects.

\begin{figure}[h]
  \centering
  \includegraphics[width=1.0\linewidth]{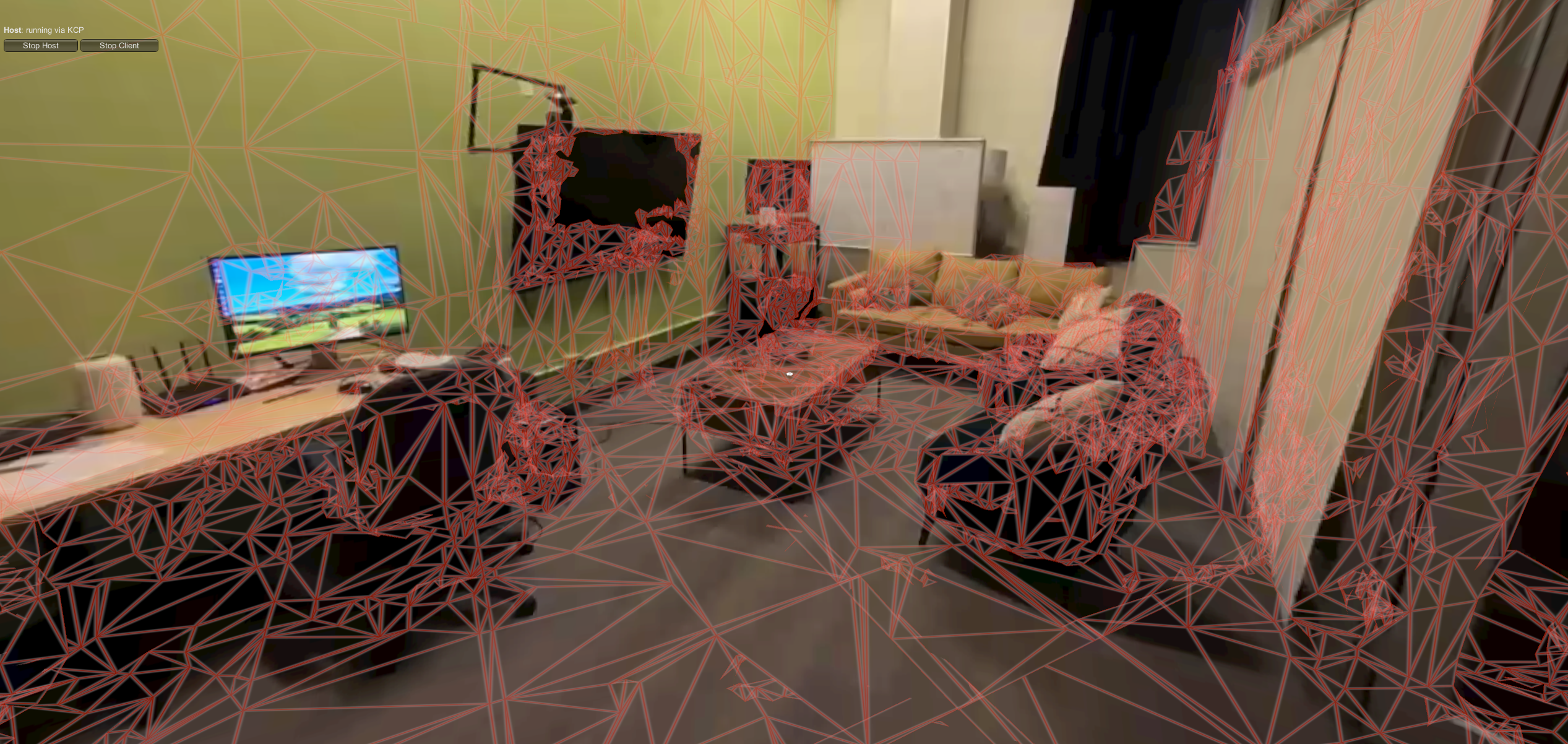}
  \caption{Spatial-accurate Alignment of 3D Reconstruction with the \TSD{} Video. The edges of the spatial mesh are only coloured in red here for demonstrative purposes.}
  \Description{We see a projected 360-degree video frame that is displaying a lab space. On the 360-degree video frame, we see a spatial mesh with edges coloured in red overlaid. The spatial mesh and the 360-degree video frame are spatial-accurately aligned. Note that the edges of the spatial mesh are invisible to the users, they are only coloured in red in this figure for demonstrative purposes.}
  \label{fig:aligned_mesh}
\end{figure} 

\subsection{Combining \TSD{} Video with Spatial Mesh}
\label{sec: spatial_mesh}
\textit{VirtualNexus} embeds a spatially aligned 3D reconstruction of the physical environment with the \TSD{} video, laying out the basis for spatial interaction with physicality. To achieve this, we utilized the spatial meshes created and maintained by Microsoft HoloLens 2, which is the foundation of a mixed-reality experience. As HoloLens creates or updates a spatial mesh, \textit{VirtualNexus} extracts the vertices and triangles from the spatial meshes and transforms them into the VR world's coordinates. \textit{VirtualNexus} then sends the mesh information through a TCP connection to the remote VR side and reconstructs the spatial meshes in real-time. To accurately align the \TSD{} video with the reconstructed spatial mesh, we reverse-engineered the \TSD{} camera's intrinsic parameters and projected the \TSD{} video to the skybox with equidistant fisheye mapping\footnote{\url{https://docs.opencv.org/3.4/db/d58/group__calib3d__fisheye.html}}. We show the alignment between the spatial meshes and the \TSD{} video in Fig. \ref{fig:aligned_mesh}. We implement spatial mesh synchronization in a silent thread to keep it seamless for both AR and VR users.

\subsection{Monocular-Binocular Trade-off}
\label{sec: mono-bino}
In virtual and augmented reality, virtual contents are rendered binocularly to generate a depth cue. However, as most \TSD{} videos are monocular, users can only tell depth using their empirical knowledge of objects' sizes (i.e., closer objects look bigger and farther objects are smaller). We started with overlaying binocularly rendered objects in front of a monocular \TSD{} video. However, we found that this causes an inconsistency regarding depth perception: a virtual object looks closer than a physical object in the \TSD{} video even if they are placed in the same position. To mitigate this, in the VR build, we shifted the position of the right-eye camera leftwards by the VR headset's inter-pupillary distance (IPD), causing the VR headset to effectively render in monocular mode, thus rendering virtual objects as if they belonged to the \TSD{} video. Such an adaptation may lead to difficulty in perceiving depth during object manipulation. In the future, we can opt to use binocular \TSD{} cameras (already available as commodity products) creating \TSD{} videos with binocular depth perception.

\subsection{Spatially Synchronized Collaboration}
\label{sec:objects}
As mentioned in \ref{sec: sys_arch}, we aligned the AR and VR space using the QR Codes attached to the \TSD{} camera as the spatial anchor. To facilitate synchronized remote collaboration, we implemented synchronized ray pointers, annotations, and virtual objects.

\subsubsection{Synchronized Ray Pointers and Annotations}
\begin{figure*}[h]
  \centering
  \includegraphics[width=0.95\linewidth]{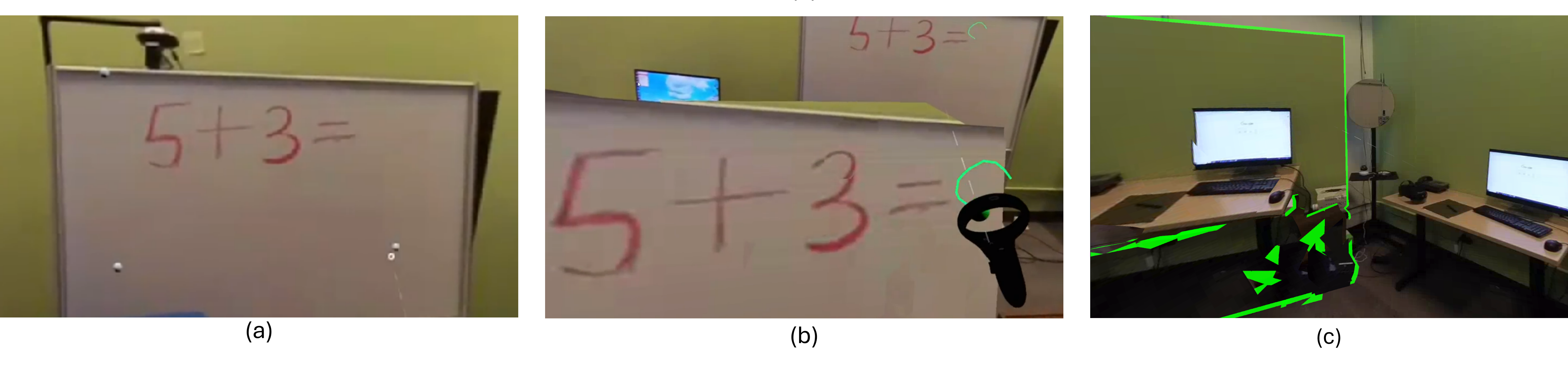}
  \caption{Environment Cutouts: In (a), the VR user defines a cutout of the whiteboard through 4 raycasted points. In (b), annotations on active cutouts are synced to the original location. In (c), users can cutout 3D space additional to 2D surfaces. }
  \Description{In (a), we see the perspective of the VR user as they move the pointer around the remote environment to define four dots on a whiteboard. These dots become the cutout shown in (b), as the remote VR user has brought the cutout closer to them in (b). As they annotate on the cutout whiteboard, we see that the original position also reflects the same annotation. In (c) we see the perspective of the VR user as they make a cutout of a desk. The desk cutout has 3D depth and is shown beside the original desk in the environment.}
  \label{fig:cutout}
\end{figure*}

It is common to use ray pointers to convey ideas and intentions in virtual and augmented applications. In \textit{VirtualNexus}, the AR and VR users can see each other's hand/controller ray pointers, which are implemented by constantly exchanging ray origin and direction information using UDP packets. We implemented shared annotations by adding two additional bytes to the same UDP packets exchanging ray pointer positions: a byte that indicates whether a user is drawing and a byte indicating the number of annotations a user has drawn. The drawing flag is set to 1 when a user is drawing annotations in their own world (the VR user presses a controller button and the AR user uses a pinch gesture), causing the user's synchronized pointer in the other user's world to draw annotations at the same time. We use the number of annotations as a sequence number to detect when a user starts a new annotation or deletes the latest annotation. We implemented the annotations with Unity line renderers. Users can create floating annotations or draw on the environment. In the latter case, we attach the annotations (as child objects) to the scene objects (e.g., depth mesh or environment cutouts) they are drawn on. To distinguish the ownership of annotations, the users see their own annotations in green and the collaborator's annotations in red. The VR user can switch between far and near annotations (i.e., between the ray pointer and the ``poke'' pointer). In the near annotation mode, a green sphere is rendered at the position of the poke pointer to indicate the status of annotation (see Fig. \ref{fig:cutout}b).

\subsubsection{Shared Virtual Objects}\label{sec: share_vo}
Both the AR and VR users can spawn shared virtual objects (see Fig. \ref{fig:app1}a). Virtual objects appear in the same location in the world for both the AR and VR user and can be freely controlled by either user via grab interactions. Users can also choose to edit their physics properties, their material, etc. Motions and edits are fully synchronized between sides by using a server-client setup built atop the Mirror\footnote{\url{https://mirror-networking.com/}} library. Our implementation initially provides a default set of meshes representing some base shapes, such as a cube, sphere, etc. The local AR user can extend this set by scanning physical objects into shared virtual replicas. We detail the virtual replica creation in Sec. \ref{sec: virtual_rep}.

\subsection{Environment Cutouts}
\label{sec:cutout}
To define an environment cutout, the VR user first makes a selection of 4 points with their ray pointer cast onto the depth mesh (Fig. \ref{fig:cutout}a). These raycast points and the camera position define a selection frustum. We then select the triangles of the spatial mesh that lie in the selection frustum, which form new mesh objects that define the cutout. While users can create both 2D (Fig. \ref{fig:cutout}b) and 3D cutouts (Fig. \ref{fig:cutout}c), the latter may be subject to occlusions.

The cutout supports standard VR manipulations, such as grabbing, rotating, and scaling. Users can ``select'' a cutout to make it active, which causes the VR user's actions to be performed relative to the cutout, and causes their avatar to be rendered in AR relative to the cutout's physical counterpart (e.g. as shown in \ref{fig:app2}(b)). With no cutout, or if the cutout is deselected, the VR user's interactions will occur with respect to the \TSD{} video, and they will be rendered at the location of the \TSD{} camera (as in Fig. \ref{fig:app2}d).

We sync interactions across this copied cutout and the world-space \TSD{} video. When the VR user creates a virtual object (annotation or mesh) while a cutout exists, they see two objects --- one corresponding to the world space (the ``original object''), and one relative to the cutout (``copy object''), for example, in Fig. \ref{fig:cutout}b. Movements and edits are synchronized between the original, copy, and the virtual objects displayed to the AR user, but the copy itself is only visible to the VR user when using a cutout. 

\subsection{Virtual Replica Creation with Instant-NGP}\label{sec: virtual_rep}

\textit{VirtualNexus} allows the AR user to create shared virtual replicas from physical objects in the environment. We chose Neural Radiance Fields (NeRF) to create virtual replicas from a futuristic standpoint: NeRFs have shown promise in producing photorealistic scans of objects and scenes with high-quality lighting and texture, and we feel that future object scanning pipelines may involve such technologies for fidelity, rather than traditional pipelines like KinectFusion \cite{Izadi2011KinectFusion}. However, for compatibility with existing mesh-rendering pipelines, we produce both a NeRF model and traditional vertex-coloured mesh from our object scanning pipeline. 

\begin{figure}[h]
  \centering
  \includegraphics[width=1.0\linewidth]{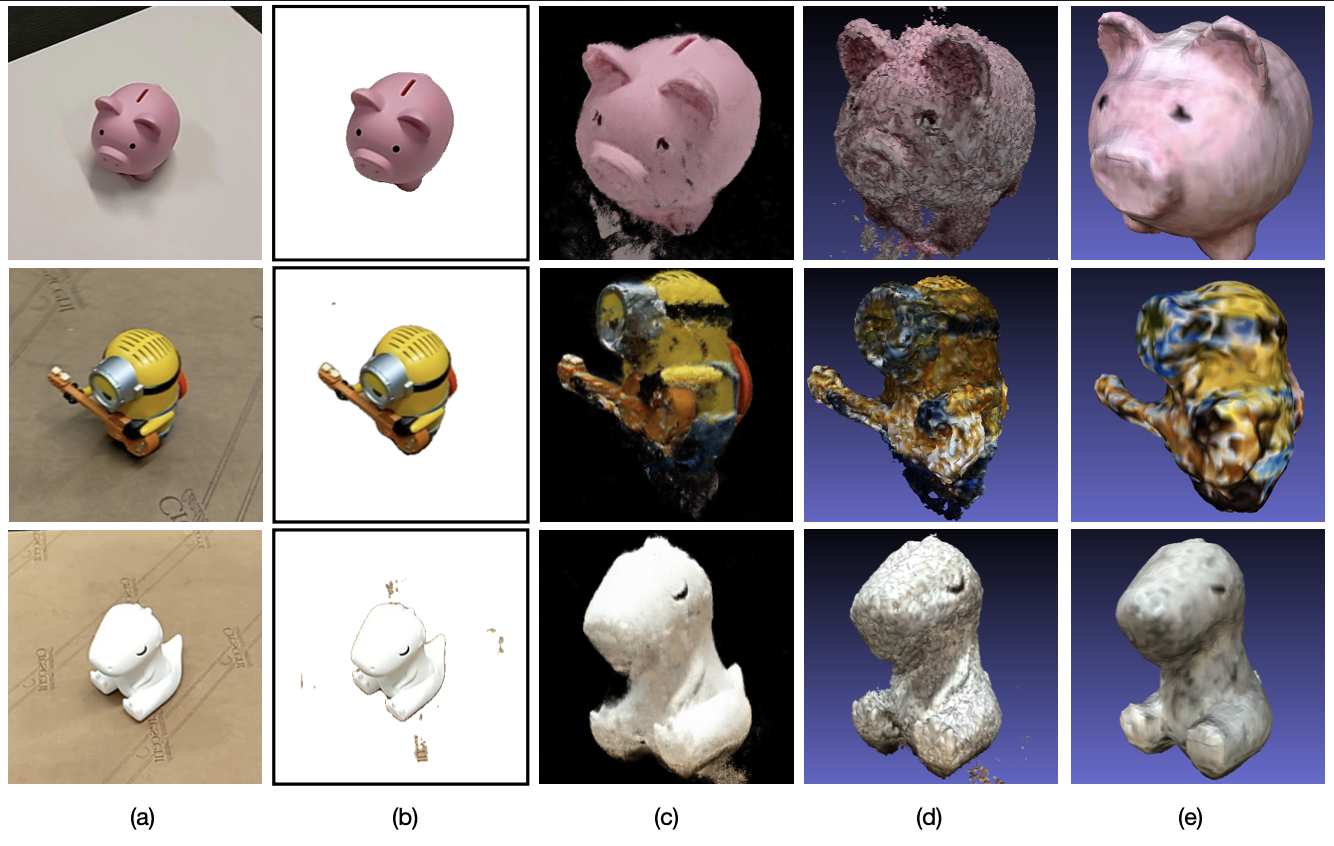}
  \caption{Intermediate results for rapid virtual replica creation: (a) original RGB image (b) background removed image (c) volume rendering by Instant-NGP (d) mesh object created from cube-marching (e) Voxelized and smoothed final object.}
  \Description{15 images presented in this figure: the first row shows a pink pig; the second row shows a yellow minion; the third row shows a white dinosaur. The first column shows one of the images from the multi-view 2D image set; the second column shows the background removed image; the third column shows Instant-NGP created volumetric rendering object; the fourth column shows the Triangulated mesh object; the fifth column shows the Blender processed final object}
  \label{fig:scan_creation}
\end{figure} 

To the best of our knowledge, our system is the first to adopt NeRF reconstruction for remote AR/VR collaboration. Among the variants of NeRF scene reconstruction techniques, Instant-NGP~\cite{mueller2022instant} strikes a balance between training time and quality. In our preliminary exploration, reconstructing individual objects with Instant-NGP (i.e., as opposed to an entire scene) with sufficient quality only takes 1--3 minutes on our NVIDIA GeForce RTX 4090 GPU machine. As we require a user to walk around the targeted object, our pipeline is best suitable for smaller desktop objects (e.g., small appliances, toys, hand-held tools). \textit{VirtualNexus}' end-to-end virtual replica creation pipeline has a server-client architecture (see bottom left of Fig. \ref{fig:sys_arch}). The server is implemented in Python and incorporates Instant-NGP's Python API~\footnote{\url{https://github.com/NVlabs/Instant-NGP}}. Examples of intermediate results at each stage of the pipeline can be seen in Fig. \ref{fig:scan_creation}.

\subsubsection{Colour and Depth Image Capturing}
To scan an object, a user triggers the function in the AR application and then walks around the target object. A semi-transparent grey rectangle is rendered to help the user center the object in their field of view (FoV). We capture colour and depth images of the object using the native Universal Windows API \footnote{\url{https://learn.microsoft.com/en-us/windows/uwp/audio-video-camera/process-media-frames-with-mediaframereader}} and HoloLens 2 Research Mode API \footnote{\url{https://github.com/microsoft/HoloLens2ForCV}} at 5 FPS for about 15 seconds, accumulating around 75 images. For each frame, we reproject the depth image from the perspective of the colour camera to align the depth and colour images, then stream the resulting images to the reconstruction server. The depth images are then used for background segmentation and improving the efficiency and quality of the NeRF reconstruction~\cite{kangle2021dsnerf}. 

\subsubsection{Pre-processing: Background Segmentation} 
To reconstruct a clean virtual replica of an object, we first remove the background, which we assume is a planar surface (e.g. a table or platform). In each image, we start with the plane obtained from the HoloLens' built-in plane detection functionality, then use a RANSAC algorithm to refine the fit~\cite{Xiao2018MRTouch}. We select all non-planar points as the initial `coarse mask' of foreground pixels. Subsequently, we obtain a refined segmentation mask from Segment-Anything~\cite{kirillov2023segment} using the average of the coarse mask as a point prompt. Segment-Anything~\cite{kirillov2023segment} outputs a hierarchy of masks, and we use the one that best overlaps with the coarse mask as the final segmenting mask. Our background segmenting process takes about 25 seconds.

\subsubsection{Colmap and Instant-NGP}
Before providing the images to Instant-NGP\cite{mueller2022instant}, we need to obtain the camera poses for the images. Initially, we tried to use the HoloLens' reported camera poses for each frame directly, but found that the poses were not accurate enough for satisfactory reconstruction. Therefore, we used Colmap~\cite{schoenberger2016mvs, schoenberger2016sfm}, a structure-from-motion technique that is used by most NeRF variants. We fed the images and the Colmap-determined camera poses to Instant-NGP, which reconstructs the object as a NeRF model and also outputs a vertex-coloured mesh with cube-marching. Running Colmap is the most expensive part of our pipeline, and can take anywhere from 20 seconds to 2 minutes. By contrast, Instant-NGP's training process takes about 15 seconds while cube-marching is practically instantaneous.

\subsubsection{Post-processing: Mesh Simplification and Smoothing}
The initial mesh created by Instant-NGP contains too many vertices and often contains unsightly holes. Therefore, we apply a ``Remesh Modifier'' with voxelization and smooth shading using the Blender API\footnote{\url{https://docs.blender.org/api/current/index.html}} on the initial mesh. This process both simplifies and smooths the mesh, and takes about 3 seconds. Our virtual replica creation server sends the mesh information as an obj file with per-vertex colours to both the AR and VR builds. We implemented a parser that can process colourized obj files at runtime, allowing either the AR or VR user to create them as shared virtual objects (Section \ref{sec: share_vo})

\section{Application Scenarios}

Here we outline \textit{VirtualNexus}' application to various domains.

\subsection{Content Authoring and Prototyping}
Collective prototyping is a key application domain for collaborative mixed reality \cite{Huang2024SurfShare, nebelingXRDirector202}. \textit{VirtualNexus} facilitates such real-time content authoring through the creation, sharing, and manipulation of virtual objects and annotations for remote users. Furthermore, the scanning and creation of instant replicas allow both users to integrate shapes beyond basic primitives, bringing in virtual objects that mimic real physical items. The cutout feature provides additional options for the remote user to interact and prototype, allowing increased precision by bringing further areas closer.

\begin{figure*}[h]
  \centering
  \includegraphics[width=0.75\linewidth]{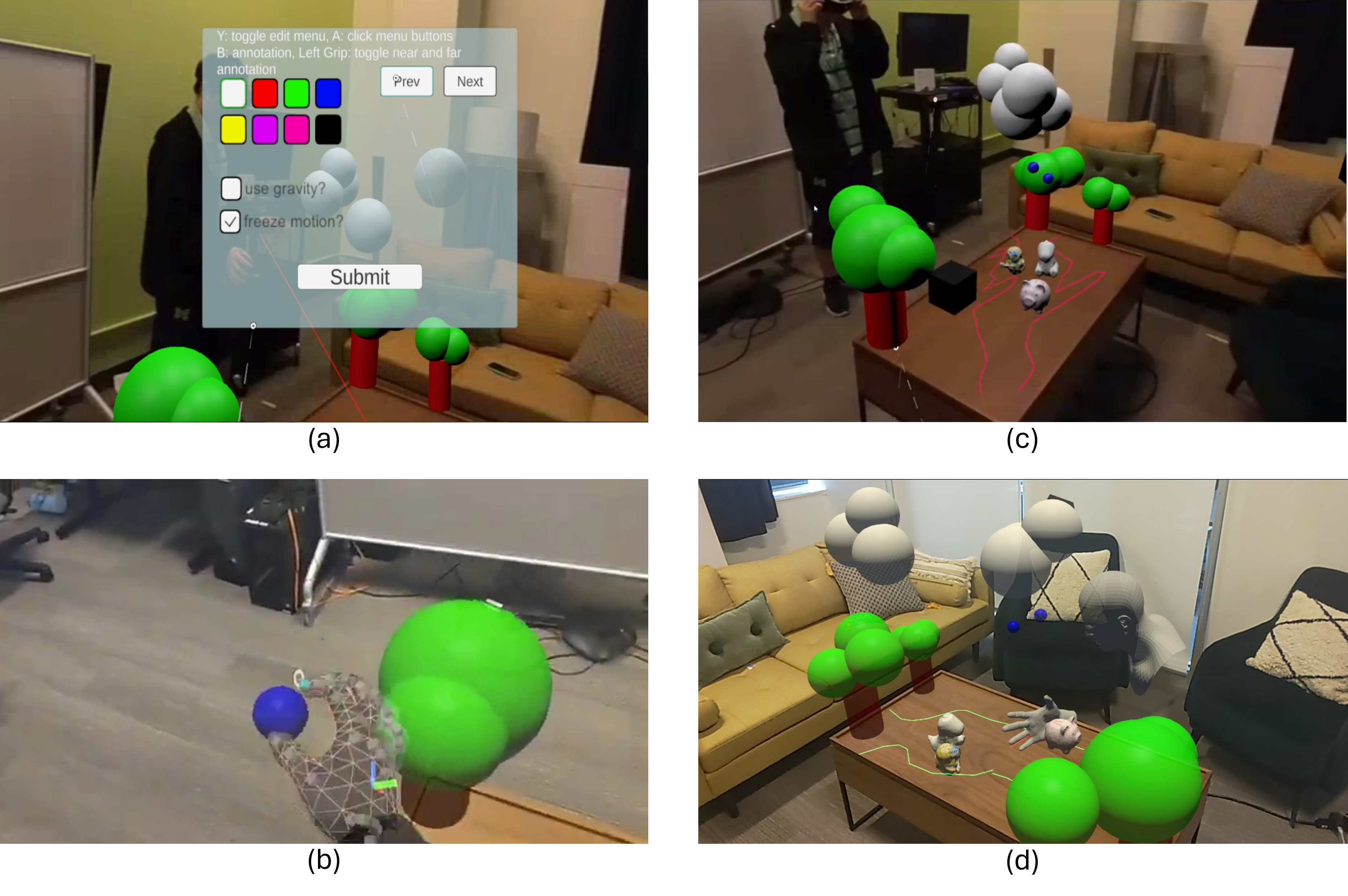}
  \caption{The collaborative prototyping scenario: The users collaboratively create a story scene. In (a) and (b), both VR and AR users use shared virtual objects for the scene. Subimages (c) and (d) show the completed scene from the VR and AR perspectives.}
  \Description{In (a), we see the VR perspective of creating a menu. There is a menu popup in the VR world that allows the user to configure the colour, the object, and the physics properties, as well as a button that says submit. In the background, we see the local AR environment and the AR user. In (b), the wee the AR perspective of manipulating a blue sphere. The AR user is moving the blue sphere around the scene, which already contains trees made of primitive shapes. In (c) and (d), we see the completed virtual scene with trees, clouds, raindrops, and characters as they walk through a forest defined on a desk.}
  \label{fig:app1}
\end{figure*}

For example, in Fig. \ref{fig:app1}, we use \textit{VirtualNexus} to create a collaborative virtual scene on a desk that the local user can walk around and view from different angles. In this scene, basic primitives such as cubes and spheres form the environment, and virtual replicas are used to create more detailed characters. Annotations are used to define areas in the environment (i.e. a path). The domain of content creation and prototyping also forms the basis of our user study (Section \ref{sec: study}), which involves collaboratively building a storyboard. 

\subsection{Remote Education and Instruction}
\begin{figure*}[h]

  \centering
  \includegraphics[width=0.75\linewidth]{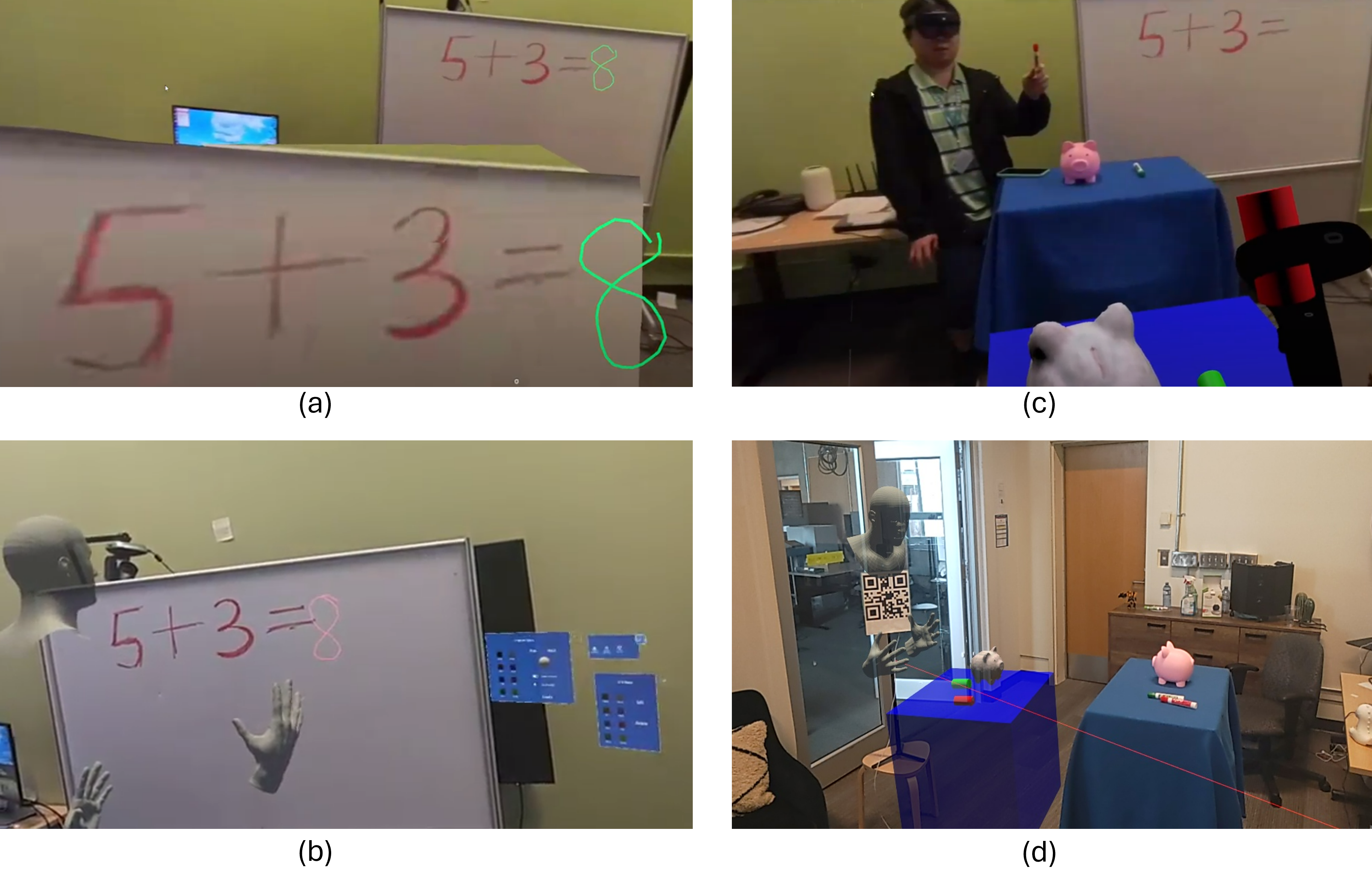}
  \caption{The remote education scenario: In (a), the VR student answers a question on the whiteboard using the pulled cutout. The AR teacher sees the answer written on the board in (b). The AR teacher also sees the VR student standing close to the board. In (c) and (d), we see an instructor and learner engaging in a crafts session. From the VR perspective in (c), virtual objects can mimic real-world items as both users hold up a red `marker'. (d) shows the AR instructor's perspective.
  }
  \Description{Subimage (a) shows the VR perspective. There is a whiteboard with the question 5+3 in the background, the VR user has made a cutout that replicates the whiteboard but is brought closer. They use this cutout to write the answer 8, which appears in green and is reflected back on the original whiteboard. In (b), the AR user sees the answer 8 on the whiteboard, as well as the VR user's avatar standing close to it. In (c), the VR user sees the AR instructor in front of a table with a model pig and markers while holding up a red marker. The VR user has a similar setup in the virtual world and is holding up a red cylinder representing the marker. Finally, in (d), the AR user sees both the original table as well as the VR learner's table, which have the same items on them.}
  \label{fig:app2}
\end{figure*}

Remote mixed reality systems also apply to remote education and instruction \cite{wuCurrentStatusOpportunities2013, kaminskaVirtualRealityIts2019, sharmaVirtualRealityClassroom2013}. Online learning platforms have become increasingly important in an increasingly digital world, and such platforms facilitate communication between educators and students despite distances \cite{cowitHowDoTeachingPractices2023}. We demonstrate a virtual classroom environment in which a teacher, in a classroom, can call upon a student, who may be joining remotely, to answer a question on a whiteboard (Fig. \ref{fig:app2}a and \ref{fig:app2}b). The teacher first annotates a question on the whiteboard. The student might find the whiteboard to be too far to interact with precisely. In real life, the student may walk up to the whiteboard. \textit{VirtualNexus} allows the student to achieve the same by bringing the whiteboard closer using its environment cutout. The student can then annotate their answer on this closer cutout, which is then reflected on the original whiteboard. 

To extend this education scenario, the teacher might ask students to mirror their interactions with virtual objects in a virtual hands-on lesson (Fig. \ref{fig:app2}c and \ref{fig:app2}d). To illustrate, we outline an arts-and-craft exercise in which the teacher is teaching about modelling and colouring while the student follows along. The teacher's desk has equipment and objects found in the classroom (i.e. markers); the student can replicate it using virtual objects. However, some required objects for the task may not be physically present for the student (i.e. the model pig). Thus, the teacher can scan the object locally and create a replica for the remote student. The student can then use these virtual objects to replicate the teacher's instructions. 

\subsection{Shared Recreational Activities}
\begin{figure*}[h]

  \centering
  \includegraphics[width=0.95\linewidth]{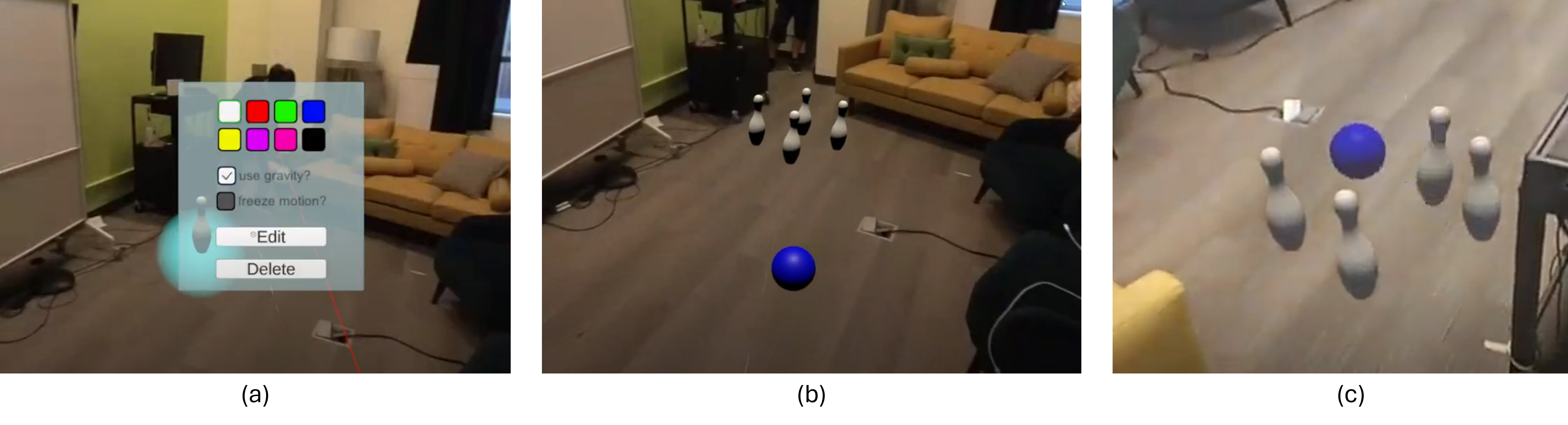}
  \caption{The shared recreational activity scenario: In (a), the remote and local users collaboratively create, edit, and arrange bowling pins with physics properties. (b) and (c) shows the users playing the remote bowling game in VR and AR.}
  \Description{Subimage (a) shows the VR perspective in the edit menu, changing the gravity of the pins. In the background, the AR user places the pins in the correct positions in the world. Subimage (b) shows a blue ball rolling towards the setup pins from the VR user's perspective; subimage (c) shows the same ball about to hit the pins but from the AR user's perspective.}
  \label{fig:app3}
\end{figure*}

Mixed reality mediums are often used in games and other recreational domains to encourage exercise and socialization. Using \textit{VirtualNexus}, we can develop collaborative recreational activities that use virtual objects for remote users. We illustrate an example using a bowling game situated on a virtual alley overlaid on the observed local environment (Fig. \ref{fig:app3}). Users can create virtual bowling pins and lay them at the end of the virtual alley. Then, either user can create a ball, which can be rolled at the pins. By taking turns rolling the balls, the users can experience a fun virtual bowling session situated in a physical environment.

\section{User Study}\label{sec: study}
We conducted a user study on \textit{VirtualNexus}. With a collaborative storyboarding task, we assessed the usability of the novel interactive techniques (e.g., environment cutouts and virtual replicas).

\subsection{Participants}
We recruited 14 participants (8 females, 5 males, averaged 24.8 years old. 1 participant reported N/A for both demographic questions) through convenience sampling, forming 7 dyads. All participants had some prior experience with remote collaborative tools (e.g. Google Docs) and video communication tools (e.g. Zoom). Almost all participants had some experience with using VR in the past (13 out of the 14 participants); experience with AR headsets was rarer (8 out of the 14 participants). Our study was reviewed and approved by the institutional ethical board, and all participants reviewed and signed a consent form prior to the study.

\subsection{Study Protocol}
Upon arrival, the participant dyad was split into a local AR user and a remote VR user. Each participant underwent an individual short guided tutorial regarding interactions using \textit{VirtualNexus} features. After the participants familiarized themselves with the system, they worked together to create a virtual storyboard for a researcher-provided narrative. This collaborative task took users through all features of the system - participants discussed and ideated the storyboard through cutouts and annotations, and then created it using virtual scanned objects. We instructed the participants to scan and create the first physical object they decided to use in the task, allowing them to experience the full virtual replica creation process. In the interest of time, we provide pre-scanned models for additional objects users may need afterwards. We continued the task until the users had finished creating enough scenes, or when we reached the allotted time. After completion, the researchers wrapped up the study through a final questionnaire. It involved 5-point Likert-scale questions that related to immersion, presence, ease of use, and task performance using the \textit{VirtualNexus} system, as well as an optional open-ended field for each question (Fig. \ref{fig:results_AR} and \ref{fig:results_VR}). The entire study took approximately 90 minutes, and participants were reimbursed \$24 CAD. 

\subsection{Results and Findings}
\begin{figure*}[h]
  \centering
  \includegraphics[width=0.98\linewidth]{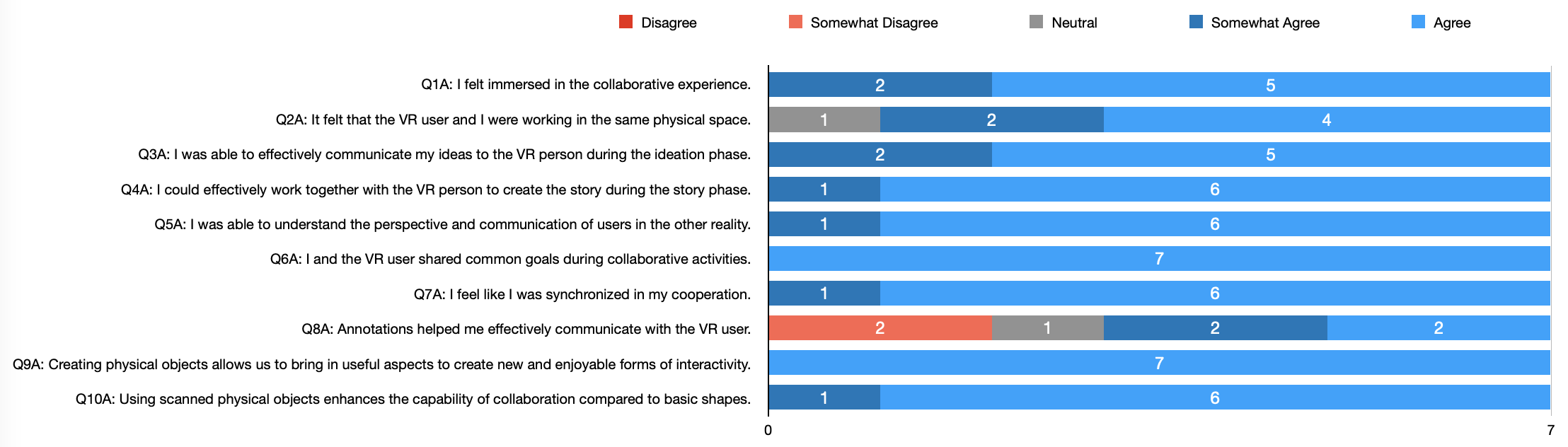}
  \caption{Results from the 5-point Likert scale questions presented to AR users.}
  \Description{Subjective 5-point Likert scale rating for AR users: The questions asked to the participants are shown on the left side of the figure including evaluation of how they think about our system features, learnability, etc. The answers are on a 5-point Likert scale. The results are shown on the right side of the figure with colour-coded correspondence to their Likert scale category.}
  \label{fig:results_AR}
\end{figure*}

\begin{figure*}[h]
  \centering
  \includegraphics[width=0.98\linewidth]{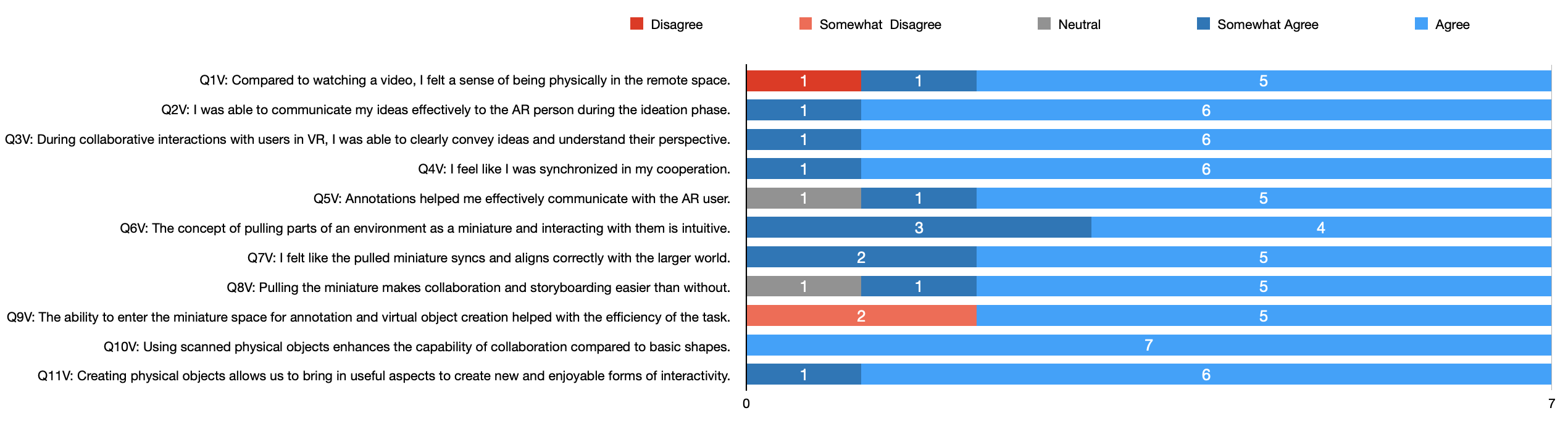}
  \caption{Results from the 5-point Likert scale questions presented to VR users.}
  \Description{Subjective 5-point Likert scale rating for VR users: The questions asked to the participants are shown on the left side of the figure including evaluation of how they think about our system features, learnability, etc. The answers are on a 5-point Likert scale. The results are shown on the right side of the figure with colour-coded correspondence to their Likert scale category.}
  \label{fig:results_VR}
\end{figure*}

Participants generally were positive towards \textit{VirtualNexus} (Fig. \ref{fig:results_AR} and \ref{fig:results_VR}). Fig. \ref{fig:teaser4} shows an example storyboard created in the study. We present the findings based on their responses to our questionnaires, especially those relevant to the two novel interactive techniques. A1-A7 and V1-V7 refer to the AR and VR users respectively.

\subsubsection{Cutout Interactions}
The VR users reported that the concept of cutting out partial environment and interacting with it was intuitive (Q6V: Mean = 4.57, STD = 0.53) --- \textit{V5: ``It was a learning curve, but it became quite intuitive after a short exposure to the experience”}. \textit{V3} echoed \textit{V5} but recommended improvements in distinguishing between the cutout and the world space (as currently, the cutout space is not localized to a smaller volume). \textit{V2} indicated that the cutouts improved clarity: i.e. \textit{``I would otherwise not be able to see what work they're doing on the whiteboard and that would make things very difficult''}. \textit{V1} also supported this approach in terms of clarity and interaction precision: \textit{``Pulling the miniature whiteboard closer was necessary for writing legibly with the annotation tool.''}. 

\subsubsection{Scanned Physical Objects}
Almost all users agreed (Q10A: Mean = 4.86, STD = 0.38; Q11V: Mean = 4.86, STD = 0.38) that having rapidly scanned physical objects enhanced the capability of collaboration compared to having only primitive shapes. From the AR perspective, both \textit{A2} and \textit{A3} thought it was more fun to interact with a virtual object compared to the same physical object. From the VR perspective, \textit{V5} mentioned that scanned objects better incorporate physical components from the scene when compared to only using the regular \TSD{} video feed. Additionally, \textit{V3} praised the usefulness of having ad-hoc created replicas in the task, they stated: \textit{“the basic shapes are not sufficient for modelling more complicated objects. (Without virtual replicas) in our task, the three characters would likely have had to be represented by geometric shapes rather than their scanned models..., potentially reducing our working efficiency”}. 

\section{Discussion and Future Work}\label{sec: discussion}
Drawing from the user study results and comparing with representative prior research, here we discuss how \textit{VirtualNexus} improves \TSD{} telepresent collaboration and identify future work. We start with the implication of embedding a 3D reconstruction under the \TSD{} video and reflect on the \textit{VirtualNexus}'s interactive techniques.

\subsection{Balancing \TSD{} and 3D reconstruction}
Past research has studied the trade-off between using \TSD{} videos and 3D reconstruction for telepresence~\cite{teoMixedRealityRemote2019a, teoTechniqueMixedReality2019a, teoExploringInteractionTechniques2020a}. In \TSD{} videos, despite the higher visual quality, remote VR users cannot move in the shared world without involving locomotive equipment such as robots \cite{JonesVROOM2021, heshmatGeocachingBeamShared2018a}. In contrast, telepresence systems with 3D reconstructions \cite{teoTechniqueMixedReality2019a, teoExploringInteractionTechniques2020a} allows users to walk around the remote environment, but the reconstructed scene suffers from holes and artifacts due to imperfect scanning and occlusion.

Teo et al. \cite{teoMixedRealityRemote2019a} set out to balance this trade-off between \TSD{} videos and 3D reconstructions in telepresence. This work allows the remote guest to switch between the \TSD{} video and the point-cloud reconstruction of the same physical environment. However, to utilize the merits of both, the user needs to frequently context switch between two distinct sets of interactive modalities, potentially increasing the mental workload. \textit{VirtualNexus} takes a slightly different approach: while the remote user always sees the physical environment in a high-quality \TSD{} video, we align a transparent 3D reconstruction with the \TSD{} video to enhance the interactivity. We believe such a design provides a more coherent interactive experience and better physical presence (Q1V from the questionnaire): In our study, we observed the remote users naturally annotating and placing virtual artifacts on remote physical surfaces, like they are physically in the remote environment wearing an AR headset.

\subsection{Interactivity in \TSD{} Video Telepresence}

In the direction of improving the interactivity of \TSD{} video telepresence, Rhee et al.~\cite{Rhee2020ARVRTeleportation} incorporated virtual annotations and artifacts. However, the main drawbacks of \TSD{} telepresence remain: The inability to walk around and interact with the remote environment.

In co-located collaboration, users can freely walk around and utilize their surroundings. Such freedom is limited for remote users telepresent with \TSD{} videos. Inspired by the concept of WiM\cite{Danyluk2021WiM}, \textit{VirtualNexus} implements the environment cutouts, which does the opposite by bringing parts of the environment toward the remote user. In our study, we found the cutouts also improved the clarity and precision: For further away areas, the remote user can pull environment cutouts closer for more precise annotation.

\textit{VirtualNexus} additionally provides the remote user with better access to individual physical objects with ad-hoc creation of virtual replicas. These objects mimic reality, but take advantage of digital affordances --- they can be replicated, scaled down or up, and can have different physics properties. Echoing past research on using virtual replicas for remote collaboration and instruction~\cite{Oda2015VirRep, Elvezio2017VirtualRep, Zhang2022gesturevirtualreplica, Huang2024SurfShare}, participants in our study found virtual replicas enhance interpretability. In contrast, representing objects as primitives adds an interpretation layer and hinders efficiency. Currently \textit{VirtualNexus} takes 1--3 mins to create a unique virtual replica. Usability research by Nielsen~\cite{Nielsen1994Usability} suggests that a response time over 10 seconds risks losing users' attention, but can be alleviated by providing a progress bar. As we managed the virtual replica creation in a separate thread behind the scenes, we expect the users can work on other sub-tasks in parallel. For example, we observed some AR participants proceed to sketch on the whiteboard with the remote VR user. Future work can reduce the processing time for object reconstruction by replacing Colmap~\cite{schoenberger2016mvs, schoenberger2016sfm} with refined HoloLens camera poses.

\subsection{Awareness of the Virtual Other and Context Switching from Cutouts}

Our study surfaced two future improvements for \textit{VirtualNexus}: 1) the local AR user's awareness of the remote VR user, and 2) context switching between the cutout space and the original environment.

AR users have more freedom to move and a smaller FoV, so they have a reduced awareness of the VR user. Such awareness is further reduced when the VR user relocates to the area they cut out from the environment. Therefore, it is reasonable to smooth out the VR user's relocation (e.g., by animating their motion) when they activate/deactivate the cutout. We can also provide additional cues to the local AR user when the VR user tries to communicate~\cite{Piumsomboon2019ShoulderGiants, Piumsomboon2017ExploreEnhancement}. 

Presently, the cutout space arbitrarily overlays the original world. The remote user needs to context switch as they redirect their focus from the cutout space to the environment and vice versa, making it hard to understand whether an object belongs to the cutout space or the original world. Future work can localize the cutout using a ``snow globe'' metaphor --- the cutout acts as a localized miniature world: Objects outside this space are hidden, and the objects within would be more easily understood as belonging to the cutout.

\section{Conclusion}
We introduce \textit{VirtualNexus}, a system for AR/VR collaboration that combines high-fidelity \TSD{} video with accurate 3D reconstructions of physical environments. It supports traditional collaborative features like annotations and virtual object manipulation and adds novel features. In VR, users can create cutouts --- miniature parts of the original world, while in AR, users can quickly scan and share physical items as virtual replicas. We outline applications of the \textit{VirtualNexus} system and perform a user study with a collaborative storyboarding task. We find that the cutout system was intuitive and provided increased clarity and precision, and the scanning system enhanced the capabilities of the collaborative processes.

\begin{acks}
This work was supported in part by the Natural Science and Engineering Research Council of Canada (NSERC) under Discovery Grant RGPIN-2019-05624 and by Rogers Communications Inc. under the Rogers-UBC Collaborative Research Grant: Augmented and Virtual Reality.
\end{acks}

\bibliographystyle{ACM-Reference-Format}
\bibliography{sample-base}

\appendix

\end{document}